\documentclass[singlecolumn,preprint,prl,aps,superscriptaddress]{revtex4-1}

\usepackage[latin9]{inputenc}
\usepackage{mathtools}
\usepackage{amsbsy}
\usepackage{amssymb}
\usepackage{amstext}
\usepackage{bm}
\usepackage{xcolor}
\usepackage{cancel}
\definecolor{darkblue}{rgb}{0, 0, 1}
\newcommand{\RN}[1]{%
	\textup{\uppercase\expandafter{\romannumeral#1}}%
}
\usepackage[unicode=true,pdfusetitle,
bookmarks=false,
breaklinks=false,pdfborder={0 0 1},backref=false,colorlinks=true,allcolors=darkblue]
{hyperref}
\hypersetup{
	bookmarksnumbered=false,bookmarksopen=false}
\makeatletter

\@ifundefined{textcolor}{}
{%
	\definecolor{BLACK}{gray}{0}
	\definecolor{WHITE}{gray}{1}
	\definecolor{RED}{rgb}{1,0,0}
	\definecolor{GREEN}{rgb}{0,1,0}
	\definecolor{BLUE}{rgb}{0,0,1}
	\definecolor{CYAN}{cmyk}{1,0,0,0}
	\definecolor{MAGENTA}{cmyk}{0,1,0,0}
	\definecolor{YELLOW}{cmyk}{0,0,1,0}
}

\newcommand{\beq}{\begin{equation}}
\newcommand{\eeq}{\end{equation}}
\newcommand{\beqa}{\begin{eqnarray}}
\newcommand{\eeqa}{\end{eqnarray}}

\usepackage{amsmath}
\usepackage{amssymb}
\usepackage{txfonts,color}
\makeatother

\begin{document}
\title{Ultrafast quantum gates with fully quantized free-electron quantum optics}

\author{Yongcheng Ding}
\affiliation{Department of Physical Chemistry, University of the Basque Country UPV/EHU, Apartado 644, 48080 Bilbao, Spain}

\date{\today}

\begin{abstract}	
Free-electron quantum optics provides a versatile platform for manipulating electrons at the quantum level with potential applications in quantum information technologies. We propose a grating-based architecture for fully quantized free-electron quantum optics, in which photon-electron interactions map onto Jaynes-Cummings and Tavis-Cummings models via Bloch-Floquet analysis. Within this framework, we design ultrafast single- and two-qubit gates with cavity-free flying electrons, enabling universal quantum computing in experimentally accessible setups. More broadly, this framework establishes a platform for probing free-electron quantum optics and advancing quantum technologies in simulation, sensing, and information processing.
\end{abstract}
\maketitle

Light-electron interaction has become central to exploring quantum effects in ultrafast electron microscopy~\cite{abajo2010optical,kirchner2014laser,dahan2020resonant,henke2021integrated} and dielectric laser acceleration (DLA)~\cite{breuer2013laser,peralta2013demonstration,adiv2021quantum}. A key milestone in this direction is photon-induced near-field electron microscopy (PINEM)~\cite{barwick2009photon,feist2015quantum,shiloh2022quantum}, which enables coherent energy exchange between electrons and optical near fields. This process allows precise manipulation of the electron wavefunction, marking the emergence of free-electron quantum optics (FEQO)~\cite{ruimy2025free}. Depending on the field strength and electron kinetic energy, distinct diffraction regimes arise, each corresponding to a different physical picture~\cite{textbook,batelaan2007illuminating,morimoto2018diffraction}. Within the FEQO framework, these regimes illustrate the wave-particle duality of free electrons, unifying the principal diffraction theory established so far. These include the Raman-Nath (fast, wave-like electron in PINEM)~\cite{park2010photon,abajo2010multiphoton,dahan2021imprinting}, Bragg (slow, wave-like electron)~\cite{eldar2024self,pan2023low,pan2024free}, Stern-Gerlach (slow electron in a gradient field)~\cite{ding2025ultrafast}, DLA (fast, particle-like electron)~\cite{gover2018dimension,pan2023weak}, and anomalous PINEM (interaction with pre-chirped electrons)~\cite{pan2019anomalous,zhou2019quantum,lin2024ultrafast}, encompassing the main manifestations of free-electron diffraction. Despite this diversity, theoretical descriptions have largely treated the optical field classically, retaining only the quantization of the electron.

Quantizing the optical field elevates the light-electron interaction from a semiclassical process to a genuinely quantum exchange, revealing phenomena inaccessible to classical fields. Within this fully quantum framework, electrons can become entangled with photons, and their wavefunctions can exhibit nonclassical features, extending FEQO beyond classical-field descriptions. In cavity-based FEQO, photon-electron interactions map onto a Jaynes-Cummings model~\cite{jaynes1963comparison,shore2007the}, supporting deterministic photon-pair generation and SWAP operations between electron and photonic qubits~\cite{karnieli2023jaynes}. These second-quantized models provide a physical platform for advanced quantum technologies, suggesting architectures in which free electrons serve as carriers of quantum information. Building on this concept, a recent proposal exploits free electrons as flying qubits, using nonlinear cavity interactions to achieve ultrafast, deterministic, and universal quantum computing~\cite{karnieli2024universal}. Meanwhile, recent works have demonstrated free-electron qubits, experimentally through temporal Talbot revivals~\cite{tsarev2021free} and theoretically via near-field encoding in even and odd sidebands~\cite{reinhardt2020free}. These studies mark key advances toward quantum control of free electrons, yet a fully quantized framework capable of describing and scaling such interactions has remained open.

Here we propose a cavity-free architecture that realizes effective Jaynes-Cummings and Tavis-Cummings model~\cite{tavis1968exact} to describe the photon-electron interaction. This approach is general across a broad range of electron velocities and field strengths, providing a versatile platform for fully quantized FEQO. This platform thus enables ultrafast quantum gates, paving way to universal free-electron quantum computing.

To study the photon-electron interaction, we consider the relativistically modified minimal coupling Hamiltonian $H = H_e + H_I$. A second-order expansion of the relativistic dispersion gives the electron kinetic energy $H_0 = E_0 + v_0(p - p_0) + (p - p_0)^2/(2\gamma^3 m_e)$, and the interaction with the classical field $A(z,t)$ reads $H_I = -(e/\gamma m_e) A \cdot p$. We also neglect the ponderomotive potential $e^2 A^2/(2\gamma m_e)$. The vector potential $A(z,t) = (E_z/\omega_L)\cos(\omega_L t - q z + \phi_0)$ describes a laser field with amplitude $E_z$, frequency $\omega_L$, and phase $\phi_0$. We define the wave vector as $q = 2\pi/\Lambda$, where $\Lambda$ is engraved on the grating to satisfy the phase-matching condition $\omega_L=v_0q$. We promote the classical field to a single-mode quantum field with photon Hamiltonian $\mathcal{H}_p = \hbar \omega_L (\hat{a}^\dag \hat{a} + 1/2)$. As the electron moves through the periodic grating, we apply Floquet-Bloch theory to expand its wavefunction as $\psi(z,t) = \sum_n c_n(t) e^{i k_n z}$, where $k_n = p_0 + n \hbar q$ and $n$ counts the photons absorbed or emitted. Meanwhile, we quantize the electron by replacing the coefficients $c_n(t)$ with annihilation operators $\hat{c}_n(t)$, which satisfy the anti-commutation relation $\{\hat{c}_n^\dag(t), \hat{c}_m(t')\} = \delta_{nm} \, \delta(t-t')$. This yields the second-quantized electron Hamiltonian $\mathcal{H}_e = \sum_n E_n  \hat{c}_n^\dag \hat{c}_n$, where the on-site energies $E_n = E_0 + n \hbar v_0 q + {n^2 \hbar^2 q^2}/(2 \gamma^3 m_e)$ produce a nonlinear energy spacing between momentum sidebands. With both light and electron quantized, the interaction becomes $\mathcal{H}_I=-eE_z\hbar/(\gamma m_e \omega_L)\sum_n[k_{n-1}\hat{c}_n^\dag\hat{c}_{n-1}\hat{a}(t) + k_{n+1}\hat{c}_n^\dag\hat{c}_{n+1}\hat{a}^\dag(t)]$. Thus, we derive the total Hamiltonian describing the light-electron interaction
\begin{eqnarray}
\label{eq:pinem}
\mathcal{H}_{\text{PINEM}} &=& \sum_n E_n \hat{c}_n^\dag \hat{c}_n +\hbar\omega_L\left(\hat{a}^\dag\hat{a}+\frac{1}{2}\right) \nonumber\\
&+& \hbar g\sum_n \left(\hat{c}_n^\dag\hat{c}_{n-1}\hat{a} + \hat{c}_n^\dag\hat{c}_{n+1}\hat{a}^\dag\right),
\end{eqnarray}
where $g=-e\tilde{E}_zk_0/(2\gamma m_e\omega_L)$ is the index-independent coupling coefficient with the approximation $k_n=k_0+nq\simeq k_0$ since $q\ll k_0$. The single-photon field amplitude $\tilde{E}_z $ is a property of the quantized mode itself, independent of the light state.

\begin{figure}
\includegraphics[width=1\linewidth]{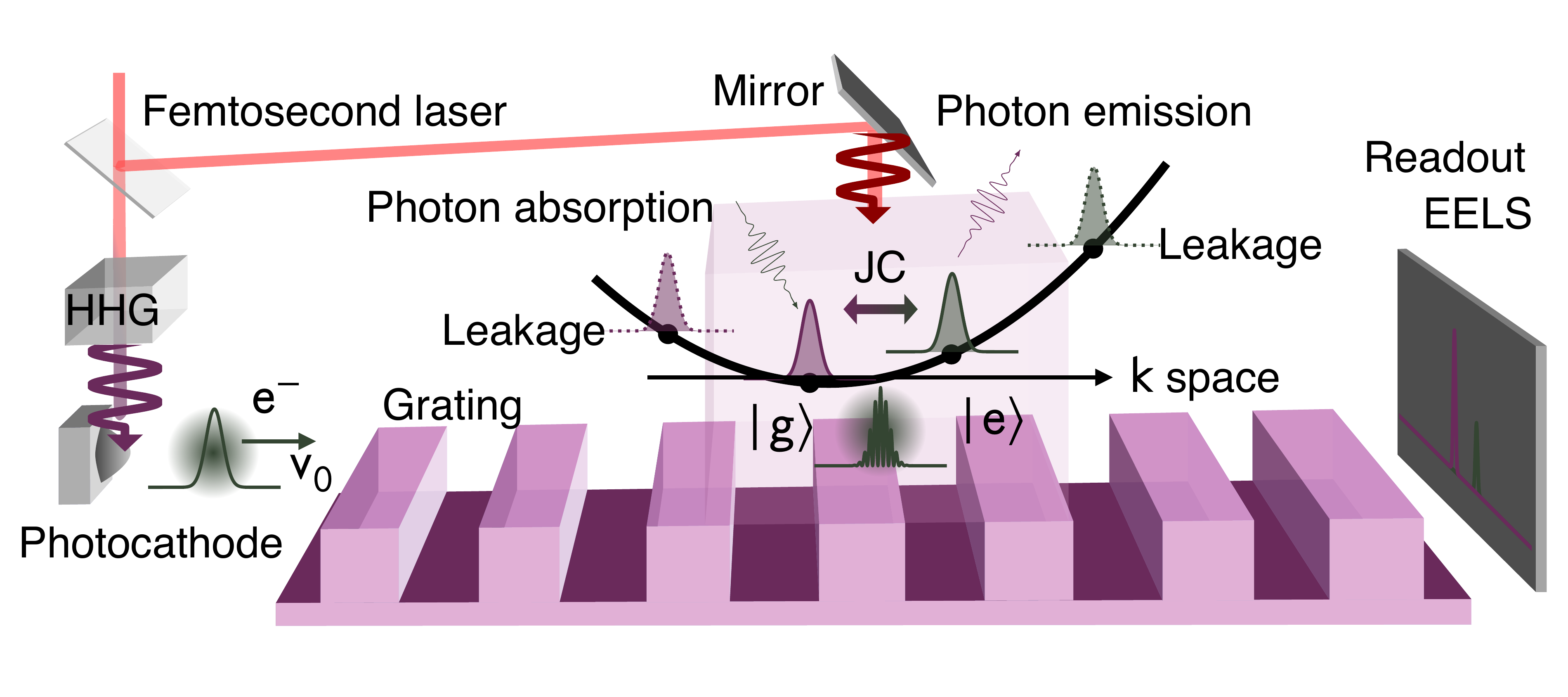}
\caption{\label{fig:scheme}Schematic setup of the photon-electron interaction realizing an effective Jaynes-Cummings model. A free electron enters the near-field interaction region, generating a light-induced synthetic dimension. The on-site energy spacing restricts the momentum sidebands to form an effective two-level electron qubit. This configuration allows coherent quantum control and the implementation of gate operations within the Jaynes-Cummings framework.}
\end{figure}

In Fig.~\ref{fig:scheme}, we illustrate the minimal experimental setup for implementing the second-quantized PINEM Hamiltonian~\eqref{eq:pinem}. We employ the same optical arrangement as in the classical case but reinterpret the modulated optical field as a quantized single mode. A femtosecond laser pulse is split into two paths. One path generates ultraviolet light via second harmonic generation to release electrons from the photocathode. The other excites a single quantized optical mode at the grating surface, described using box quantization. When the electron velocity is low, the dispersion curvature becomes significant and the interaction enters the Bragg regime, where only two momentum sidebands $n = \pm 1/2$ participate, leading to an effective Jaynes-Cummings Hamiltonian
\begin{equation}
\label{eq:JC}
H_{\text{JC}} = \frac{\hbar v_0q}{2} \hat{\sigma}_z + \hbar\omega_L\hat{a}^\dag\hat{a} + \hbar g(\hat{\sigma}_+\hat{a} + \hat{\sigma}_-\hat{a}^\dag),
\end{equation}
where $\hat{\sigma}_z = \hat{c}_{1/2}^\dag \hat{c}_{1/2} - \hat{c}_{-1/2}^\dag \hat{c}_{-1/2}$, $\hat{\sigma}_+ = \hat{c}_{1/2}^\dag \hat{c}_{-1/2}$, and $\hat{\sigma}_- = \hat{c}_{-1/2}^\dag \hat{c}_{1/2}$ are the Pauli matrices.  The derivation of effective JC Hamiltonian~\eqref{eq:JC} enables the study of quantum optics phenomena with ultrafast free electron systems, e.g., collapse and revival, deterministic photon generation, etc. When the phase-matching condition is satisfied, the interaction is resonant and should realize genuine ultrastrong~\cite{diaz2019ultrastrong} or even deep strong coupling regimes~\cite{casanova2010deep} of the JC model rather than the quantum Rabi model without rotating-wave approximation. However, at sufficiently strong coupling, the two-level approximation between the $n = \pm 1/2$ sidebands ceases to hold, as higher momentum states participate in the dynamics, leading to a multi-level extension of the JC picture. Furthermore, we generalize the Hamiltonian~\eqref{eq:JC} to $N$ free electrons, obtaining the Tavis-Cummings Hamiltonian
\begin{equation}
\label{eq:TC}
H_{\text{TC}} = \sum_i\frac{\hbar v_0q}{2} \hat{\sigma}_z^{(i)} + \hbar\omega_L\hat{a}^\dag\hat{a} +\sum_{i} \hbar g(\hat{\sigma}_+^{(i)}\hat{a} + \hat{\sigma}_-^{(i)}\hat{a}^\dag),
\end{equation}
assuming that the electrons are sufficiently far apart to neglect Coulomb interactions. This effective Hamiltonian allows the exploration of the dynamical burst of quantum light known as superradiance, as well as non-equilibrium steady states in driven-dissipative setups.

After deriving the light-electron interaction models, we construct a native gate set for universal quantum computing in the language of quantum optics. In particular, we propose implementing the Rx, Ry, and iSWAP gates using the effective JC~\eqref{eq:JC} and TC~\eqref{eq:TC} Hamiltonians as a minimal set. To preserve gate fidelity, the quantum light must disentangle from the electron wavefunction after each operation, which requires operating the PINEM system in different regimes.

For single-qubit rotation gates, we operate in the resonant regime, ensured by the phase-matching condition $v_0 q = \omega_L$. In the interaction picture, the JC Hamiltonian reads
\begin{equation}
\label{eq:JCint}
H_\text{JC}^{(\text{int})} = \hbar g (\hat{\sigma}_+ \hat{a} + \hat{\sigma}_- \hat{a}^\dag),
\end{equation}
and governs the evolution of the electron-photon state according to $i \hbar\partial_t |\psi_\text{int}\rangle = H_\text{JC}^{(\text{int})} |\psi_\text{int}\rangle.$ In practice, a resonant femtosecond laser pulse excites this quantized mode into a coherent state $|\alpha\rangle$. If the electron qubit is initialized to $|e\rangle=|1/2\rangle$, this leads to electron dynamics described by $P_e = \exp(-|\alpha|^2)\sum_m(\alpha^{2m}/m!)\cos^2(g\sqrt{m+1}t)$, exhibiting the characteristic collapse and revival behavior. We perform the single-qubit gate within the collapse envelope. For photon number $\langle \hat{a}^\dag\hat{a}\rangle = |\alpha|^2 \gg 1$, we can approximate $\hat a \to \alpha$ in the interaction Hamiltonian~\eqref{eq:JCint}, yielding an effective Rabi frequency $\Omega = 2g \sqrt{ m + 1} \approx 2g |\alpha|$. Accordingly, to implement an $\text{R}_{\text{x}}(\theta)$ gate, we set the gate duration to $T_\theta = \theta / \Omega = \theta / (2g |\alpha|)$, which is much shorter than the characteristic collapse time $t_c \sim 2\pi |\alpha| / g$, ensuring that the electron evolves nearly coherently and enabling high-fidelity rotations. Correspondingly, an $\text{R}_{\text{y}}(\theta)$ rotation requires an additional phase shift of $\pi/2$ on the light field. This control provides sufficient freedom to realize arbitrary rotations on the Bloch sphere. The standard $\text{R}_{\text{z}}(\theta)$ gate can be implemented using composite pulses $\text{R}_{\text{x}}(\pi/2)\text{R}_{\text{y}}(\theta)\text{R}_{\text{x}}(-\pi/2)$, which introduce only the additional duration of a $\pi$-pulse, avoiding the need to operate in the dispersive regime. Alternatively, one can perform a virtual $\text{R}_{\text{z}}(\theta)$ rotation by adjusting the phase of the driving field.

\begin{figure}
\includegraphics[width=1\linewidth]{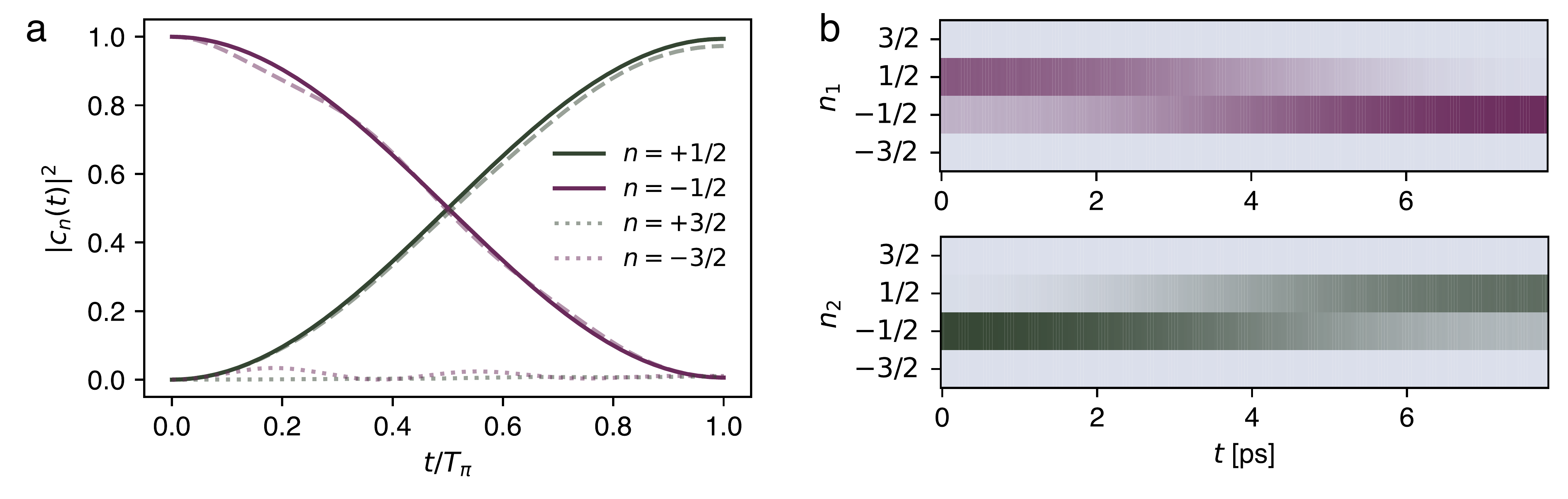}
\caption{\label{fig:gate}Numerical simulation of populations on momentum sidebands.  (a) A resonant $\pi$-pulse flips the population $|g\rangle\rightarrow|e\rangle$ of a low-energy electron ($E_0 = 100~\mathrm{eV}$) using coherent light $|\alpha|=10$ in the UV regime ($\hbar\omega_L=6.20~\mathrm{eV}$).  The dynamics closely follow the ideal Jaynes-Cummings model, with a strong vacuum field amplitude $\tilde{E}_z=7.48\times 10^6~\mathrm{V/m}$ at the diffraction limit, achieving an ultrafast gate time $T_\pi=43.3~\mathrm{fs}$ and fidelity of 0.994. The two-level approximation remains valid for slow electrons, with negligible leakage, as indicated by the dashed and dotted curves, even under a hundredfold increase of $\tilde{E}_z$. (b) Dispersive iSWAP gate mediated by a slightly detuned virtual photon of energy $6.24~\mathrm{eV}$ exchanges the populations $|eg\rangle \leftrightarrow |ge\rangle$ with a relative phase, following the Tavis-Cummings dynamics. The operation completes in $T_{\text{iSWAP}} = 7.81~\mathrm{ps}$ with $\tilde{E}_z=7.58\times 10^6~\mathrm{V/m}$, achieving a fidelity of 0.991.}
\end{figure}

Having established the theoretical framework for single-qubit rotations, we now illustrate the operation of an ultrafast $X$ gate through numerical simulation of the electron-photon dynamics governed by Eq.~\eqref{eq:pinem}, compared to the ideal JC Hamiltonian~\eqref{eq:JCint}. Figure~\ref{fig:gate}a demonstrates the evolution of populations on the electron momentum sidebands, including the computational basis states $|e\rangle = |n=1/2\rangle$ and $|g\rangle = |n=-1/2\rangle$, as well as leakage to higher and lower levels. We consider a slow electron with speed $\beta = 0.02c$ and a center energy of 100 eV, interacting with a single-mode light field of frequency 9.42 PHz in the deep ultraviolet regime, corresponding to a wave length of 200 nm and a photon energy of 6.2 eV. The grating structure has a period $\Lambda = 2\pi/q = 4$~nm, ensuring the phase-matching condition for resonant interaction. The single-photon field amplitude $\tilde{E}_z$, defined as the vacuum-mode intensity, satisfies $E_z = \tilde{E}_z\langle\hat{a}\rangle$ for a semiclassical coherent field. It depends on the quantization volume as $\tilde{E}_z = \sqrt{\hbar\omega_L/(2\varepsilon_0 V_L)}$, where $V_L$ is the box volume. We consider a box with edge length of $\lambda/2 = 100~\mathrm{nm}$, corresponding to a diffraction-limited near-field amplitude of $\tilde{E}_z = 7.48\times10^6~\mathrm{V/m}$. It leads to a weak coupling ratio $g/\omega_L = 3.85\times10^{-4}$. The gate time $T_\pi = 43.3~\mathrm{fs}$ is ultrafast, with a high fidelity of $F = \left[\mathrm{Tr}\!\left(\sqrt{\sqrt{\rho_e}\,|e\rangle\!\langle e|\,\sqrt{\rho_e}}\right)\right]^2 = 0.994$, where $\rho_e$ is the reduced electron density matrix after tracing out the light field. The quantum dynamics perfectly follow the ideal JC prediction. The entanglement entropy, defined as $S = -\mathrm{Tr}(\rho_e \ln \rho_e)$, resulting in $S/\ln 2 = 0.05$ at the end of the gate operation. This low value indicates that the qubit wavefunction remains only weakly entangled with the light field. It confirms that a coherent state with $|\alpha| = 10$ suffices to implement a resonant single-qubit rotation gate. To verify the validity of the two-level truncation under slow-electron curvature, we intentionally consider an extremely strong field of $\tilde{E}_z = 5\times10^8~\mathrm{V/m}$, corresponding to a quantization box far below the diffraction limit. the qubit approximation remains accurate, showing negligible leakage to other sidebands and achieving a fidelity of $0.973$ within a gate time of $T_\pi = 0.647~\mathrm{fs}$. Thus, we confirm that the two-level truncation remains valid for all for a wide range of physically attainable light fields, including extreme cases, with a slow electron velocity of $\beta = 0.02$.

With the single-qubit gate operations proposed, we extend our framework to two-qubit entangling dynamics. The effective TC Hamiltonian~\eqref{eq:TC} enables coherent excitation exchange between two electrons coupled to the same quantized mode, realizing the iSWAP gate. In this case, we operate in the dispersive regime $\Delta = |v_0 q - \omega_L| \gg g$, rather than at exact resonance. While the resonant TC interaction allows coherent excitation exchange, it depends on the photon number and thus fails to derive a perfect, state-independent swap for arbitrary electron states. On the contrary, in the dispersive limit, the quantized mode can be adiabatically eliminated via Schrieffer-Wolff transformation, leading to an effective coupling in the interaction picture
\begin{equation}
H_{\text{TC}}^{\text{int}} = \hbar J (\hat{\sigma}_+^{(1)}\hat{\sigma}_-^{(2)} + \hat{\sigma}_-^{(1)}\hat{\sigma}_+^{(2)})
\end{equation}
 with strength $J = g^2/\Delta$, which offers a high-fidelity iSWAP gate within gate time $T_{\text{iSWAP}}=\pi/(2J)=\pi\Delta/(2g^2)$. The electrons propagate simultaneously, while their transverse separation ensures they remain distinguishable and Coulomb interactions can be neglected.

In Fig.~\ref{fig:gate}b, we simulate the iSWAP gate in the dispersive regime by generalizing the PINEM Hamiltonian~\eqref{eq:pinem} to two electrons. Keeping the grating period for phase matching unchanged, we slightly detune the light-electron interaction by illuminating the grating structure with a laser frequency of $\omega_L = 9.49~\mathrm{PHz}$. The corresponding single-photon field amplitude for the new box size is $\tilde{E}_z = 7.58\times10^6~\mathrm{V/m}$, yielding $|g/\Delta| = 0.055$, which satisfies the condition for the dispersive regime.  The population on the momentum sidebands of each electron qubit is perfectly exchanged after $T_{\text{iSWAP}} = 7.81~\mathrm{ps}$, roughly 176 times longer than the resonant single-qubit X gate. The first and second electron qubits are initialized in the separable state $|\psi_1\rangle \otimes |\psi_2\rangle$, with $|\psi_i\rangle = \cos(\theta_i/2)|e_i\rangle + \sin(\theta_i/2)|g_i\rangle$ and angles $\theta_1 = \pi/3$, $\theta_2 = 11\pi/12$, while the light field remains in the vacuum. After applying a virtual local $Z$ rotation $U_Z = \mathrm{diag}[\exp(-i|J|T_{\text{iSWAP}}), \exp(i|J|T_{\text{iSWAP}})]$, the fidelity reaches $F = \left[\mathrm{Tr}\!\left(\sqrt{\sqrt{\rho_e}\tilde{\rho}_e\sqrt{\rho_e}}\right)\right]^2 = 0.991$, where $\tilde{\rho}_e$ is the electron density matrix after the ideal iSWAP operation.

\begin{figure}
\includegraphics[width=1\linewidth]{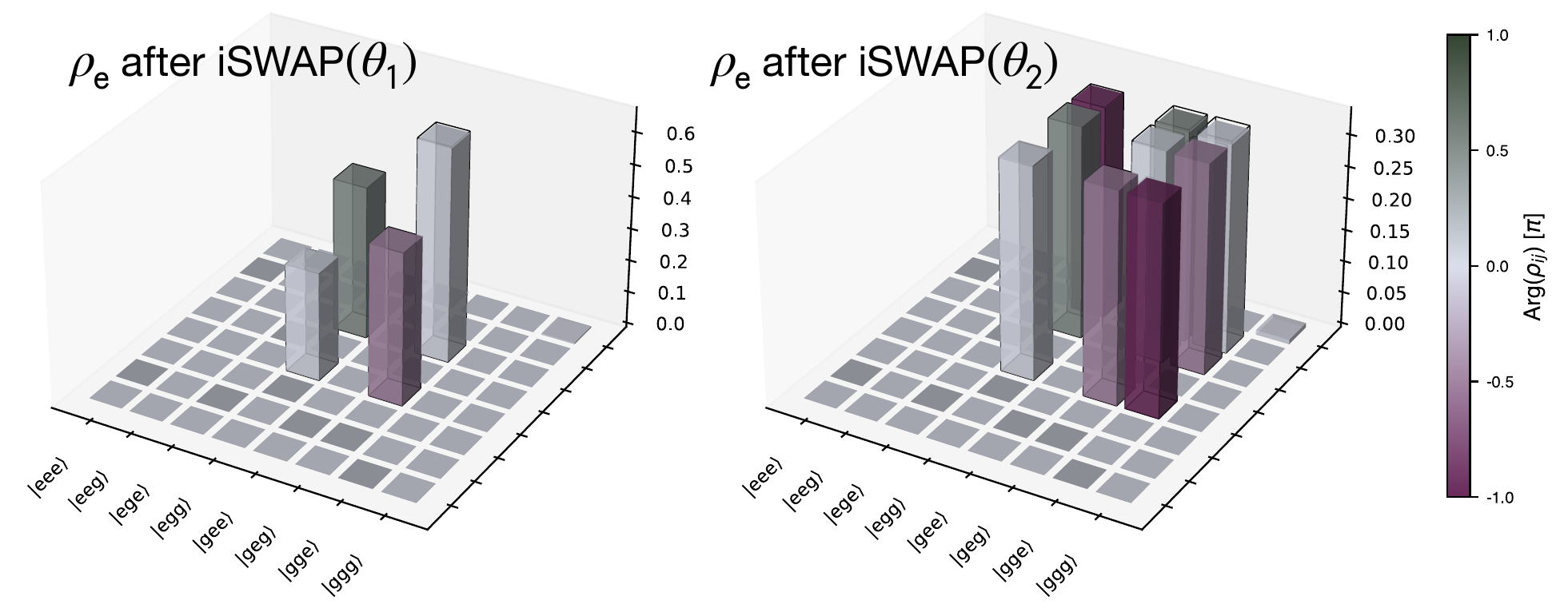}
\caption{\label{fig:W} Density matrices of a three-qubit state after applying the first and second $\mathrm{iSWAP}(\theta_k)$ gates, targeting a $W$-state. The system starts from $|egg\rangle$ with $\theta_1=\arcsin(1/\sqrt{3})$ and $\theta_2=\pi/4$. 
Ideal matrix elements are shown as transparent boxes. The ultrafast gate sequence prepares the $W$-state with a fidelity of 0.992 within $8.65~\mathrm{ps}$. A relative phase on the second qubit can be removed by a virtual $\mathrm{R_z}(\pi/2)$ rotation to obtain a standard $W$-state.}
\end{figure}

Building on the realization of the native gates, we demonstrate the preparation of a $W$ state, highlighting the universality of our light-electron quantum control scheme. The iSWAP gate used here operates in the dispersive regime, where the interaction time can be tuned as $T_{\theta} = \theta\Delta / (2g^2)$ to implement a partial $\mathrm{iSWAP}(\theta)$. Starting from the $N$-qubit initial state $|eg\cdots gg\rangle$, we sequentially apply $\mathrm{iSWAP}(\theta_k)$ between the $k$th and $(k+1)$th qubits, with $\theta_k = \arcsin\!\left({1}/{\sqrt{N - k + 1}}\right)$. This choice of angles ensures that the single-excitation amplitudes on all qubits become equal, $1/\sqrt{N}$, realizing the target $W$ state. For illustration, Fig.~\ref{fig:W} shows the electron density matrix after each $\mathrm{iSWAP}(\theta_k)$ in the preparation of a three-qubit $W$ state.  We initialize the electron state in $|egg\rangle$ with the light field in the vacuum mode. The rotation angles are $\theta_1=\arcsin(1/\sqrt{3})$ and $\theta_2=\pi/4$, corresponding to gate durations $T_{\theta_1}=4.75~\mathrm{ps}$ and $T_{\theta_2}=3.90~\mathrm{ps}$, respectively, while all other parameters remain identical to those used for the dispersive $\mathrm{iSWAP}$ simulation. The fidelities after each gate are $F_1=0.996$ and $F_2=0.992$. The target state acquires a relative phase $-i/\sqrt{3}$ on the basis $|geg\rangle$ due to the intrinsic phase of the $\mathrm{iSWAP}$ operation, which can be removed by an additional $\mathrm{R_z}(\pi/2)$ rotation on the second qubit to recover the standard $W$ state. While the sequential $\mathrm{iSWAP}(\theta)$ sequence provides a digital route to $N$-qubit $W$-state generation, the same platform offers a faster, deterministic realization through a resonant TC interaction~\eqref{eq:TC}. When all electron qubits initially in $|g\cdots g\rangle$ couple resonantly to a single-photon Fock state $|n=1\rangle$, the photon excitation transfers coherently to the symmetric bright state within $T_{\mathrm{TC}}=\pi/(2g\sqrt{N})$. This one-shot process requires only collective coupling and timing control, forming a native analog Hamiltonian block that fits naturally into a digital-analog quantum computing framework~\cite{parra2020digital}. The TC interaction thus offers a compact and high-fidelity path to entanglement generation, achieving, for example, a 3-qubit $W$ state within $250~\mathrm{fs}$ in our free-electron quantum computing architecture.

To scale the scheme up, multiple electrons can fly along the grating, transversely separated to suppress Coulomb interactions. The laser, controlled in amplitude, phase, and timing, interacts only with selected electrons to implement single- and two-qubit gates. This provides parallel or sequential operations, forming a scalable, flying-electron quantum computing architecture. Qubit states can be measured by detecting the electron momentum sidebands along the flight direction using EELS positioned perpendicular to the $Z$ axis. We measure qubit states by detecting the electron momentum sidebands along the flight direction using EELS positioned perpendicular to Z direction. We transversely displace electrons to resolve each qubit individually, while the momentum sidebands encode the logical state. With typical ultrafast PINEM parameters, momentum shifts of a few meV and transverse separations of tens of nanometers allow us to distinguish multiple qubits simultaneously.

Beyond scaling up and readout, the essential principles and practical aspects of coherent electron-light interactions merit discussion. We describe the zero-point field using box quantization, formally identical to conventional mode quantization in free space. In our setup, the mode forms a phase-matched Bloch field along the grating, removing the "flying-out-of-the-box" limitation while preserving the Hamiltonian and coupling constants used in the main text. Periodic nanostructures, such as gratings or photonic crystals, can support these Bloch modes with controllable frequency, wave vector, and polarization. In particular, the first grating harmonic $m=1$ in the quantum Smith-Purcell process~\cite{smith1953visible} provides the momentum exchange necessary to generate the same momentum sidebands that encode the qubit states, corresponding to a grating period $\Lambda=4.08~\mathrm{nm}$. The small correction relative to the classical phase-matching value leaves the preceding derivation valid to leading order. Another consideration is whether the quantized optical field on a grating can be described by box quantization. A more rigorous treatment may require mesoscopic quantum electrodynamics~\cite{rivera2020light}, since simple box quantization can be insufficient for temporally near-field modes coupled to an open nanostructure. In practice, the validity of the quantization model can be tested by observing vacuum Rabi oscillations of a single electron interacting with the mode. This also provides a means to calibrate parameters and benchmark gate operations before scaling to multiple qubits.

We have derived the fully quantized light-electron interaction theory for a PINEM system, reproducing Jaynes-Cummings and Tavis-Cummings models under well-justified approximations. This framework enables ultrafast quantum-gate design in experimentally accessible setups, advancing toward universal quantum computing with slowly flying free electrons. We demonstrated high-fidelity single- and two-qubit gates in resonant and dispersive regimes, and achieved robust state preparation by combining these gates. Future extensions could implement larger multi-qubit gates and more complex quantum operations with flying electrons. Gate design may benefit from advanced quantum control techniques to improve robustness and fidelity beyond simple resonant and dispersive gates. Coupling these capabilities with nanophotonic structures or programmable gratings allows precise control over Bloch-mode properties, enabling scalable quantum computing platforms and applications in quantum simulation, hybrid electron-photon architectures, and more complex quantum algorithms. Throughout this work we have focused on the idealized coherent dynamics governed by the quantized PINEM Hamiltonian. Incorporating dissipation, photon leakage, or electron decoherence would require an open system treatment, which lies beyond the present scope but represents an important direction for future studies. Moreover, extending this framework to quantum sensing~\cite{ruimy2021toward,gorlach2021ultrafast,karnieli2023quantum} and imaging~\cite{priebe2017attosecond,asban2021generation,bucher2024coherently,gaida2024attosecond} could open new avenues for probing matter in the fully quantum realm.

\section*{Acknowledgement} 
Y. D. thanks the European Commission for a Marie Curie PF grant (No. 101204580 FELQO). Discussions with Bin Zhang, Jorge Cassanova, and Yiming Pan are appreciated.

\pagebreak
\widetext
\begin{center}
	\textbf{ \large Supplemental Material: Ultrafast quantum gates with fully quantized free-electron quantum optics}
\end{center}

\setcounter{equation}{0} \setcounter{figure}{0} \setcounter{table}{0}
\setcounter{page}{1} \makeatletter \global\long\def\theequation{S\arabic{equation}}
\global\long\def\thefigure{S\arabic{figure}}
\global\long\def\bibnumfmt#1{[S#1]}
\global\long\def\citenumfont#1{S#1}

\section{Second quantization of mininal coupling Hamiltonian}
We start from the one-dimensional Schr\"odinger equation of the free electron interacting with light field minimal coupling
\begin{equation}
i\hbar\partial_t\Psi(z,t) = (H_0 + H_p + H_I)\Psi(z,t),
\end{equation}
where $H_0 = E_0 + v_0(p - p_0) + (p - p_0)^2/(2\gamma^3 m_e)$ is the kinetic energy of the free electron expanded to second-order with the relativistic dispersion, $H_p$ is the energy of near-field on the grating structure, and $H_I$ is the minimal coupling term. Different from the classical field Hamiltonian in the main text, here the wave function describes the photon-electron state.

In classical theory, the light field Hamiltonian reads 
\begin{equation}
H_p = \frac{1}{2} \int dr \left[\varepsilon_0E^2(r,t)+\frac{1}{\mu_0}B^2(r,t)\right],
\end{equation}
where $E(r,t) = -\partial_tA(r,t)$ and $B(r,t)=\nabla\times A(r,t)$. Considering a classical vector potential $A(z,t) = (E_z/\omega_L)\cos(\omega_L t - q z)$, we can write down the corresponding operator after second quantization as
\begin{equation}
\hat{A}(z,t) = \frac{\tilde{E}_z}{2\omega_L}\left[\hat{a}(t)e^{iqz} + \hat{a}^\dag(t)e^{-iqz}\right],
\end{equation}
where $q=2\pi/\Lambda$ is the light-induced wave vector depending on engraved grating period $\Lambda$ and $\tilde{E}_z = \sqrt{\hbar\omega_L/(2\varepsilon_0 V_L)}$. Here $V_L$ is the box size for renormalization to ensure the single-photon energy is $\hbar\omega_L$ inside the box. We employ classical phase matching condition of the Smith-Purcell type interaction, that the group velocity equals the phase velocity $v_0 q= \omega_L$, therefore, we get the quantized Hamiltonian for photons through the vector potential operator $\hat{A}(z)$
\begin{equation}
\mathcal{H}_p = \hbar \omega_L (\hat{a}^\dag \hat{a} + 1/2).
\end{equation}
The free electron traveling on the periodic grating structure. In this way, we can employ the Floquet-Bloch theory to expand the electron wave function into Bloch modes
\begin{equation}
\psi(z,t) = \sum_n c_n(t)e^{ik_nz},
\end{equation}
where $k_n=k_0+nq$ gives the synthetic dimension in $k$-space by absorbing and emitting photons.
To second quantize the electron kinetic energy, we replace the wave function and coefficient by field operator and annihilation operator that satisfie the anti-commutation relations $\{\hat{\psi}^\dag(z,t),\hat{\psi}(z,t)\}=\delta(z-z')\delta(t-t')$ and $\{\hat{c}_n^\dag(t), \hat{c}_m(t')\} = \delta_{nm} \, \delta(t-t')$, respectively. Thus, we have
\begin{eqnarray}
\mathcal{H}_e &=&~\int dz\hat{\psi}^\dag(z,t) H_e(p) \hat{\psi}(z,t)\nonumber\\
&=& \int dz~\hat{c}_m^\dag\hat{c}_n\left[e^{-ik_mz}H_e\left(-i\hbar\partial_z\right)e^{ik_nz}\right]\nonumber\\
&=& \int dz~\hat{c}_m^\dag\hat{c}_ne^{i(k_n-k_m)}\left[E_0+v_0(\hbar k_n-p_0) +\frac{(\hbar k_n-p_0)^2}{2\gamma m_e}\right].
\end{eqnarray}
The integral gives the quantized electron Hamiltonian
\begin{equation}
\mathcal{H}_e = \sum_n E_n\hat{c}_n^\dag\hat{c}_n,
\end{equation}
where on-site energy reads
\begin{equation}
E_n = E_0 + n\hbar v_0q + \frac{n^2\hbar^2q^2}{2\gamma m_e}.
\end{equation}
Accordingly, by expanding the minimal coupling Hamiltionian $(p-eA)^2/(2m_e)$ and neglecting the ponderomotive tern $\sim A^2$, we have the interaction Hamiltonian $H_I=-e/(\gamma m_e)A\cdot p$. With quantization of both the electron Hamiltonian and the near-field Hamiltonian, we derive the second-quantized interaction Hamiltonian
\begin{eqnarray}
\mathcal{H}_I &=& -\frac{e}{\gamma m_e} \int dz~\hat{\psi}^\dag(z,t)\hat{A}\cdot\hat{p}\hat{\psi}(z,t)\nonumber\\
&=& -\frac{e\tilde{E}_z}{2\gamma m_e\omega_L} \sum_{mn} \hat{c}_m^\dag\hat{c}_n\int dz [\hat{a}(t)e^{iqz} + \hat{a}^\dag(t)e^{-iqz}](e^{-ik_mz}\hat{p}e^{-ik_nz})\nonumber\\
&=&  -\frac{e\hbar\tilde{E}_z}{2\gamma m_e\omega_L} \sum_{mn} \hat{c}_m^\dag\hat{c}_n k_n \int dz~[\hat{a}(t)e^{i(k_n-k_m+q)z} + \hat{a}^\dag(t)e^{i(k_n-k_m-q)z}]\nonumber\\
&=& -\frac{e\hbar\tilde{E}_z}{2\gamma m_e\omega_L} \sum_{mn} \hat{c}_m^\dag\hat{c}_n k_n [\hat{a}(t)\delta(k_n-k_m+q) + \hat{a}^\dag(t)\delta(k_n-k_m-q)]\nonumber\\
&=& -\frac{e\hbar\tilde{E}_z}{2\gamma m_e\omega_L} \sum_n[k_{n-1}\hat{c}_n^\dag\hat{c}_{n-1}\hat{a}(t) + k_{n+1}\hat{c}_n^\dag\hat{c}_{n+1}\hat{a}^\dag(t)]
\end{eqnarray}
Considering $q\ll k_n = k_0+nq$, we approximate the coupling constant by assuming $k_n\approx k_0$
\begin{equation}
g=-\frac{e\tilde{E}_zk_0}{2\gamma m_e\omega_L}.
\end{equation}
Taken together, we have the total Hamiltonian
\begin{equation}
\label{sqe:pinem}
\mathcal{H}_{\text{PINEM}} = \sum_n E_n \hat{c}_n^\dag \hat{c}_n +\hbar\omega_L\left(\hat{a}^\dag\hat{a}+\frac{1}{2}\right) + \hbar g\sum_n \left(\hat{c}_n^\dag\hat{c}_{n-1}\hat{a} + \hat{c}_n^\dag\hat{c}_{n+1}\hat{a}^\dag\right),
\end{equation}
which is valid for any speed of the free electron. Similar to the classical case, this fully-quantized Hamiltonian also enters the Bragg regime when the electron is slow enough, i.e., only two neighbouring momentum sideband are involved in the photon-electron interaction. We have the effective Jaynes-Cummings Hamiltonian by letting $n=\pm 1/2$
\begin{equation}
\label{seq:JC}
H_{\text{JC}} = \frac{\hbar v_0q}{2} \hat{\sigma}_z + \hbar\omega_L\hat{a}^\dag\hat{a} + \hbar g(\hat{\sigma}_+\hat{a} + \hat{\sigma}_-\hat{a}^\dag),
\end{equation}
where $\hat{\sigma}_z = \hat{c}_{1/2}^\dag \hat{c}_{1/2} - \hat{c}_{-1/2}^\dag \hat{c}_{-1/2}$, $\hat{\sigma}_+ = \hat{c}_{1/2}^\dag \hat{c}_{-1/2}$, and $\hat{\sigma}_- = \hat{c}_{-1/2}^\dag \hat{c}_{1/2}$ are the Pauli matrices. Note that with phase-matching condition satisfied, the JC model is in resonant regime.

\section{Collapse and Revival, Bragg regime, and Raman-Nath regime}
In the main text we demonstrated the construction of resonant single-qubit gate in the Bragg regime within the collapse envelop. Here we go for more details of the PINEM Hamiltonian~\eqref{sqe:pinem}, to see the corresponding diffraction regimes in fully-quantized theory of photon-electron interaction.

For illustrating the collapse and revival, the initial photon state is a coherent state $|\alpha\rangle=e^{-\frac{|\alpha|^2}{2}}\sum_m\frac{\alpha^m}{\sqrt{m!}}|m\rangle$. Assuming that the electon is in $|g\rangle=|-1/2\rangle$, then we have the population on the excited state $|e\rangle=|1/2\rangle$ as
\begin{eqnarray}
\label{eq:Pe_def}
P_e(t) = |\langle 1/2|\Psi(t)\rangle|^2 &=& e^{-|\alpha|^2}\sum_m\frac{\alpha^{2m}}{m!}\sin^2(g\sqrt{m+1}t).
\end{eqnarray}
We denote $\bar n\equiv|\alpha|^2$ as the mean photon number and 
$P_m=e^{-\bar n}\bar n^{m}/m!$ as the corresponding Poisson distribution. Using $\sin^2x=\frac12(1-\cos2x)$, Eq.~(\ref{eq:Pe_def}) can be rewritten as
\begin{equation}
P_e(t)=\frac{1}{2}\Big[1-S(t)\Big],\qquad S(t)=e^{-\bar n}\sum_{m=0}^{\infty}\frac{\bar n^m}{m!}\cos\!\big(2g t\sqrt{m+1}\big).
\label{eq:Pe_sum}
\end{equation}
For large $\bar n$, the photon-number distribution is sharply peaked around $m=\bar n$ with width $\sqrt{\bar n}$.  
We introduce the deviation $x=m-\bar n$ and expand the Rabi frequency
\begin{equation}
2g\sqrt{m+1} \approx 2g\sqrt{\bar n+1}+\frac{g}{\sqrt{\bar n+1}}x-\frac{g}{4(\bar n+1)^{3/2}}x^2+\cdots.
\label{eq:expansion}
\end{equation}
Substituting this into Eq.~(\ref{eq:Pe_sum}) and replacing the Poisson distribution by a Gaussian
\begin{equation}
P_m \approx \frac{1}{\sqrt{2\pi\bar n}}\exp\!\Big[-\frac{(m-\bar n)^2}{2\bar n}\Big],
\end{equation}
the discrete sum can be approximated by an integral
\begin{equation}
\mathcal{S}(t)\equiv e^{-\bar n}\!\sum_m\!\frac{\bar n^m}{m!}e^{i\,2g t\sqrt{m+1}}\;\approx\;e^{i\theta_0(t)}\!\!\int_{-\infty}^{\infty}\!\frac{dx}{\sqrt{2\pi\bar n}}\,\exp\!\Big[-\frac{x^2}{2\bar n}+iA(t)x+iB(t)x^2\Big],
\label{eq:Gaussian_integral}
\end{equation}
where $\theta_0(t)=2g t\sqrt{\bar n+1}$,
$A(t)=g t/\sqrt{\bar n+1}$,
and $B(t)=-g t/[4(\bar n+1)^{3/2}]$.
The observable in Eq.~(\ref{eq:Pe_sum}) is $S(t)=\Re\mathcal{S}(t)$.

Neglecting the quadratic term $B(t)$, the integral in Eq.~(\ref{eq:Gaussian_integral}) becomes the characteristic function of a Gaussian, yielding
\begin{equation}
\mathcal{S}(t)\approx e^{i\theta_0(t)}\exp\!\Big[-\tfrac{1}{2}\bar n A(t)^2\Big]\simeq e^{i\,2g t\sqrt{\bar n+1}}\exp\!\Big[-\tfrac{1}{2}g^2 t^2\Big].
\end{equation}
Hence,
\begin{equation}
S(t)\approx e^{-\,\frac{g^2 t^2}{2}}\cos\!\big(2g t\sqrt{\bar n+1}\big),
\end{equation}
and the excited-state population becomes
\begin{equation}
P_e(t)\approx \frac{1}{2}\Big[1 - e^{-\,\frac{g^2 t^2}{2}}\cos\!\big(2g t\sqrt{\bar n+1}\big)\Big].
\label{eq:collapse_form}
\end{equation}
Equation~(\ref{eq:collapse_form}) describes rapid Rabi oscillations at the mean frequency $2g\sqrt{\bar n+1}$, modulated by a Gaussian envelope $e^{-g^2 t^2/2}$ that characterizes the collapse. Defining the collapse time $t_{\mathrm{coll}}$ by $e^{-g^2 t_{\mathrm{coll}}^2/2}=e^{-1}$ gives
\begin{equation}
t_{\mathrm{coll}}\approx \frac{\sqrt{2}}{g}.
\end{equation}
The physical origin of the collapse is the dephasing among the frequency components $g\sqrt{m+1}$ associated with different photon numbers $m$. The coherent superposition of many slightly detuned Rabi oscillations causes destructive interference on the timescale $t_{\mathrm{coll}}\sim 1/g$.

The higher-order (quadratic) term $B(t)x^2$ in Eq.~(\ref{eq:expansion}) becomes relevant at longer times and leads to rephasing of the dephased components. A simple estimate of the revival time follows by requiring the phase difference between adjacent photon-number components to be an integer multiple of $2\pi$:
\begin{equation}
2g t_{\mathrm{rev}}\!\left(\!\sqrt{m+2}-\sqrt{m+1}\!\right)\!
\approx \frac{g t_{\mathrm{rev}}}{\sqrt{\bar n+1}}=2\pi,
\end{equation}
which gives
\begin{equation}
t_{\mathrm{rev}}\simeq \frac{2\pi\sqrt{\bar n+1}}{g}\;\approx\; \frac{2\pi\sqrt{\bar n}}{g}.
\end{equation}
At $t=t_{\mathrm{rev}}$ the discrete phase spectrum realigns, restoring the oscillations of $P_e(t)$, this is the revival. Because $t_{\mathrm{rev}}\propto\sqrt{\bar n}$ while $t_{\mathrm{coll}}\sim 1/g$, the two timescales are well separated for large $\bar n$.  

For self-consistency with the collapse time in the main text, we examine the phase difference between adjacent photon-number components after time $t$:
\begin{align}
\Delta\phi_{\mathrm{adj}}(t) 
&= \big[\Omega_{m+1} - \Omega_m \big] t
= g\big(\sqrt{m+2} - \sqrt{m+1}\big)t.
\end{align}
For large photon number, we expand the square roots
\begin{equation}
\sqrt{m+2} - \sqrt{m+1} \simeq \frac{1}{2\sqrt{m+1}}  \simeq \frac{1}{2|\alpha|}.
\end{equation}
Hence, the typical phase difference between adjacent photon-number terms becomes
\begin{equation}
\Delta\phi_{\mathrm{adj}}(t) \simeq \frac{g t}{2|\alpha|}.
\label{eq:phase_diff}
\end{equation}
The oscillations of $P_e(t)$ collapse when the phase spread between successive photon-number components becomes sufficiently large that their contributions interfere destructively. 
A convenient criterion for this condition is that the phase difference between adjacent components reaches $\pi$, i.e.
\begin{equation}
\Delta\phi_{\mathrm{adj}}(t_c) = \pi.
\end{equation}
Therefore the collapse time is
\begin{equation}
t_c = \frac{2\pi|\alpha|}{g}.
\label{eq:tcoll}
\end{equation}
This means that after a time $t_c$, the Rabi oscillations associated with neighboring photon numbers differ in phase by $\pi$, leading to near-complete dephasing of the total signal. 

\begin{figure}
\includegraphics[width=1\linewidth]{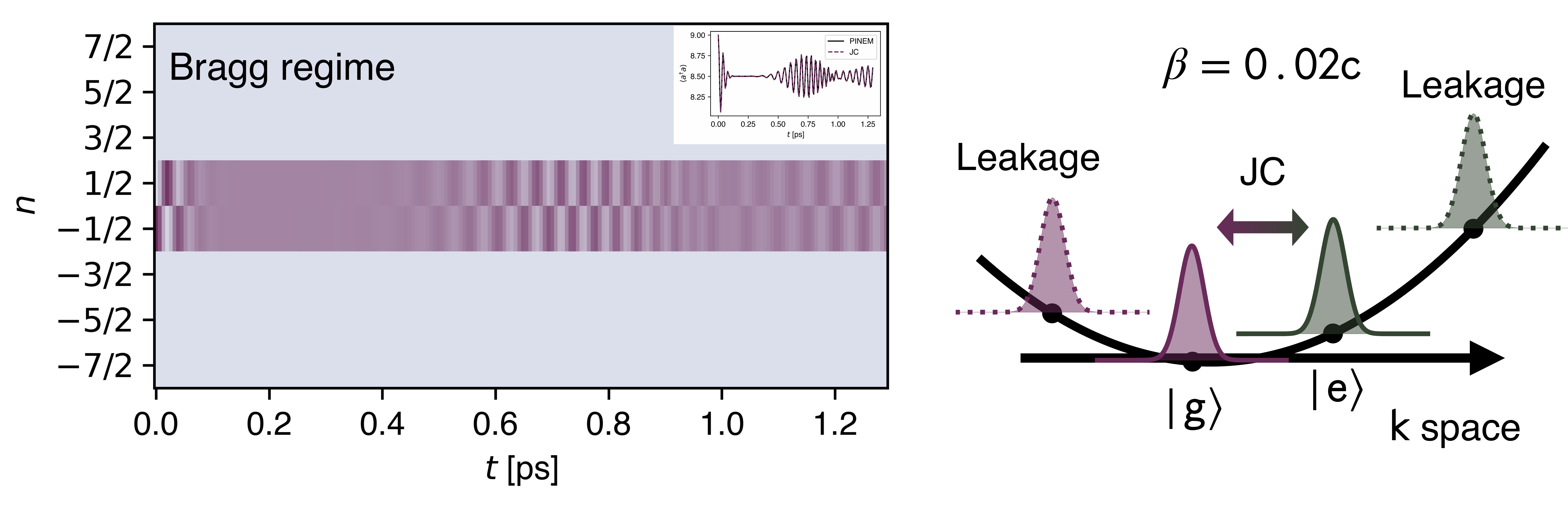}
\caption{\label{sfig:bragg} Quantum dynamics showing collapse and revival in the Bragg regime. The dispersion curvature of the slow electron ($\beta = 0.02$) suppresses higher and lower momentum sidebands, ensuring dynamics that follow the Jaynes-Cummings model exactly.
}
\end{figure}

Here we demonstrate the collapse and revival dynamics in the Bragg regime. As discussed in the main text, the two-level approximation remains valid even under an extreme vacuum field amplitude of $\tilde{E}_z = 5\times10^8~\mathrm{V/m}$ and a coherent field with $\alpha = 10$ for a slow electron with velocity $\beta = 0.02c$. For illustrative purposes, we show in Fig.~\ref{sfig:bragg} the dynamics governed by the PINEM Hamiltonian~\eqref{sqe:pinem} for a smaller coherent amplitude $\alpha = 3$ over a simulation time of $T = 1290~\mathrm{fs}$. Other parameters are fixed as $\omega_L = v_0 q = 9.42~\mathrm{PHz}$ and $g = 0.24~\mathrm{PHz}$. The PINEM dynamics show good agreement with the Jaynes-Cummings (JC) dynamics~\eqref{seq:JC}, confirming that the dispersion curvature effectively enforces the two-level truncation of the momentum sidebands. The characteristic collapse time~\eqref{eq:tcoll} is estimated as $t_c = 77.6~\mathrm{fs}$.

\begin{figure}
\includegraphics[width=1\linewidth]{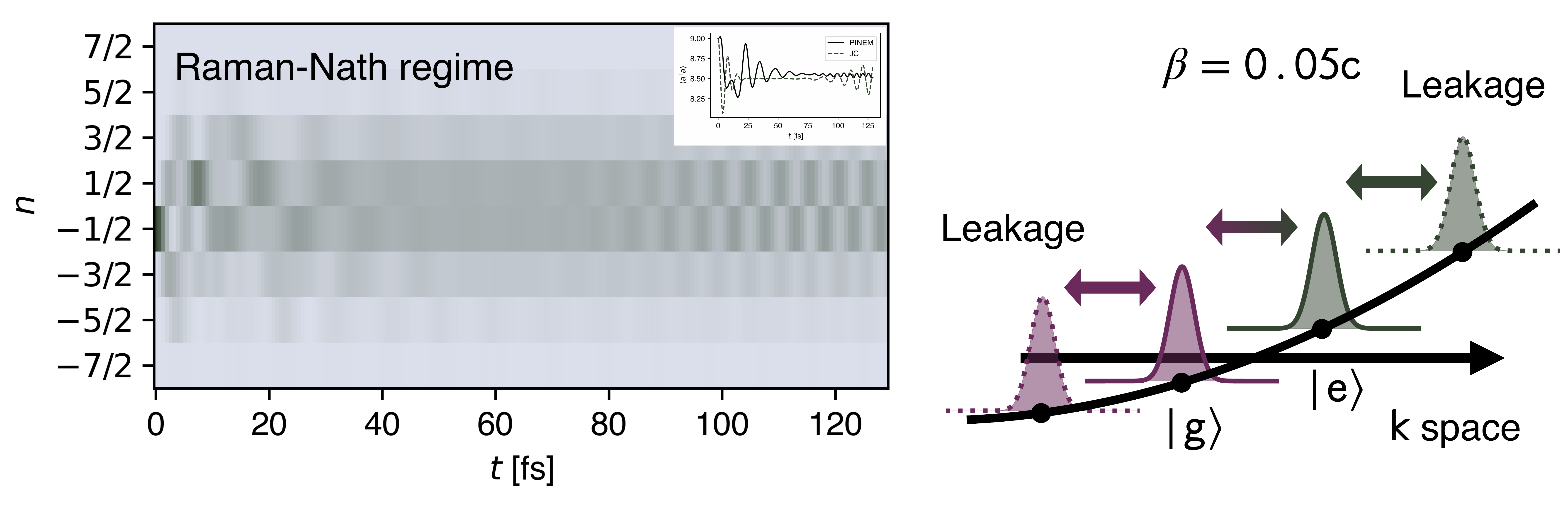}
\caption{\label{sfig:raman-nath} Quantum dynamics showing collapse and revival in the Raman-Nath regime. For the faster electron ($\beta = 0.05$), the dispersion curvature no longer suppresses the higher and lower momentum sidebands, leading to collapse-revival behavior characteristic of a multilevel system rather than the standard Jaynes-Cummings or quantum Rabi model in the ultrastrong coupling regime without the RWA.}
\end{figure}

With the same setup, we enter the Raman-Nath regime by increasing the electron velocity to $\beta = 0.05c$ and amplifying the field to $\tilde{E}_z = 10^9~\mathrm{V/m}$. Accordingly, the coupling strength becomes $g = 1.21~\mathrm{PHz}$, and the ratio $g / \omega_L > 0.1$ places the system in the ultrastrong coupling regime. In Fig.~\ref{sfig:raman-nath}, we show the corresponding PINEM dynamics within a simulation time of $T = 129~\mathrm{fs}$. Unlike the Bragg case, the dynamics no longer follow the Jaynes-Cummings model, as the two-level approximation breaks down due to the nearly linear dispersion curvature. Population leakage to higher-order sidebands such as $|\pm 5/2\rangle$ becomes visible.

\section{Quantum correction for Smith-Purcell interaction}

The momentum exchange between the electron and the photon in the grating mediated Smith-Purcell interaction obeys a discrete phase-matching (momentum conservation) condition along the electron trajectory,
\begin{equation}
k_{\parallel}^{(\mathrm{ph})} + m G = \frac{\omega_L}{v_0},
\label{eq:quantumPM}
\end{equation}
where $k_{\parallel}^{(\mathrm{ph})} = k_{\mathrm{ph}}\cos\theta$ is the photon momentum component parallel to the surface ($k_{\mathrm{ph}} = 2\pi / \lambda$), $G = 2\pi / \Lambda$ is the grating reciprocal lattice vector, $\omega_L$ the photon angular frequency, and $v_0$ the electron velocity. The integer $m$ labels the grating harmonic that mediates the interaction.

In the classical Smith-Purcell configuration, the interaction is usually described in terms of the continuous phase-matching condition we used in the main text
\begin{equation}
\omega_L = v_0 q,
\end{equation}
where $q = 2\pi / \Lambda$ is the effective spatial frequency set by the grating. This relation ensures that the electromagnetic field "chases'' the moving electron, so that their relative phase remains stationary during interaction. When the light is slightly detuned, the phase matching breaks down, and the interaction strength drops sharply.

In the quantum realm, the grating cannot be regarded as a continuous medium. Its spatial periodicity provides discrete momentum quanta $\pm m G$ that can compensate the mismatch between the photon and electron. Equation~(\ref{eq:quantumPM}) thus generalizes the classical condition by including this discrete reciprocal lattice momentum. The integer $m$ corresponds to the order of the grating harmonic participating in the interaction. Importantly, $m = 0$ yields no coupling, since it corresponds to a uniform surface with no spatial modulation. The grating must supply nonzero momentum to exchange energy between the photon and the free electron.

The value of $m$ is determined experimentally by the combination of  photon wavelength $\lambda$, incidence angle $\theta$, electron velocity $v_0$,  and grating period $\Lambda$. From Eq.~(\ref{eq:quantumPM}), one can solve for any of these parameters to achieve resonance for a desired $m$. For normal incidence ($\theta = 0$), this gives
\begin{equation}
\Lambda = \frac{2\pi m}{\dfrac{\omega}{v_0} - k_{\mathrm{ph}}}.
\label{eq:grating_period}
\end{equation}
For the parameters used in the main text, $\lambda = 200~\mathrm{nm}$ and $v_0 = 0.02c$, the classical phase-matching condition ($m=0$) yields $\Lambda = 4.00~\mathrm{nm}$.
In contrast, for the first grating harmonic ($m=1$),  Eq.~(\ref{eq:grating_period}) givess $\Lambda = 4.08~\mathrm{nm}$, which corresponds to the first quantum correction of the Smith-Purcell-type interaction. This small deviation highlights the discrete nature of the momentum exchange.

In summary, in the grating-based configuration, the wave number $q$ that defines the synthetic momentum transfer arises from the grating periodicity, with $q=2\pi/\Lambda$ for the classical (m = 0) case. For the quantum correction, the discrete grating harmonics contribute additional reciprocal lattice vectors $mG$, $G=2\pi/\Lambda$, leading to an effective longitudinal momentum transfer $q_{\text{eff}}=k_{\parallel}^{(\mathrm{ph})} + m G$.

\end{document}